\documentclass[aps,showpacs]{revtex4}
\usepackage{amssymb}
\usepackage[dvips]{graphicx}
\usepackage[english]{babel}
\usepackage{indentfirst}
\usepackage{amsxtra}
\usepackage{amsmath}
\usepackage{bbold}
\newcommand{\beq}{\begin{equation}}
\newcommand{\eeq}{\end{equation}}
\newcommand{\beqn}{\begin{eqnarray}}
\newcommand{\eeqn}{\end{eqnarray}}

\newcommand{\bsigma}{\mbox{\boldmath $\sigma$}}
\newcommand{\btau}{\mbox{\boldmath $\tau$}}
\newcommand{\half}{\frac{1}{2}}

\newcommand{\br}{{\bf r}}

\newcommand{\caI}{$^{40}$Ca\,\,}
\newcommand{\caII}{$^{48}$Ca\,\,}
\newcommand{\zr}{$^{90}$Zr\,\,}
\newcommand{\pb}{$^{208}$Pb\,\,}
\newcommand{\caIIp}{$^{48}$Ca}
\newcommand{\zrp}{$^{90}$Zr}
\newcommand{\pbp}{$^{208}$Pb}
\newcommand{\bn}{\overline{n}}
\newcommand{\bpp}{\overline{p}}
\newcommand{\bpI}{\overline{p}'}
\newcommand{\bnI}{\overline{n}'}
\newcommand{\tp}{T$_+$\,\,}
\newcommand{\tm}{T$_-$\,\,}

\newcommand{\hfp}{{\cal Q}}

\newcommand{\hdb}{{\cal B}}

\newcommand{\tup}{\tau_+}
\newcommand{\tdo}{\tau_-}

\newcommand{\dms}{\Delta_{\rm SD}}
\newcommand{\dfer}{\Delta_{\rm F}}
\newcommand{\dias}{\Delta_{\rm F}}
\newcommand{\dgt}{\Delta_{\rm GT}}
\newcommand{\drnp}{r_n-r_p}

\usepackage[usenames]{color}
\usepackage{ulem}

\begin{document}

\noindent
\title{Charge-exchange excitations with finite range interactions including tensor terms}

\author{V. De Donno}
\affiliation{Dipartimento di Matematica e Fisica ``E. De Giorgi'',
  Universit\`a del Salento, I-73100 Lecce, ITALY}
\author{G. Co'}
\affiliation{Dipartimento di Matematica e Fisica ``E. De Giorgi'',
  Universit\`a del Salento and, 
 INFN Sezione di Lecce, I-73100 Lecce, ITALY}
\author{M. Anguiano, A. M. Lallena}
\affiliation{Departamento de F\'\i sica At\'omica, Molecular y
  Nuclear, Universidad de Granada, E-18071 Granada, SPAIN}
\date{\today}

\bigskip
\begin{abstract}
We study charge-exchange excitations in doubly magic-nuclei by using a
self-consistent Hartree-Fock plus Random Phase Approximation model.
We use four Gogny-like finite-range interactions, two of them containing
tensor forces. We investigate the effects of the various parts of the
tensor forces in the two computational steps of our model, and  we find
that their presence 
is not negligible and improves the agreement with
the experimental data.
\end{abstract}

\bigskip
\bigskip
\bigskip


\maketitle
\section{Introduction}
\label{sec:intro}
The study of charge-exchange excitations in neutron-rich nuclei is an
important issue not only for its intrinsic interest
\cite{ost92,ich06,fuj11}, but also for the
special role that these excitations play in many astrophysical
processes such as beta decay, electron capture and r-process in
nucleosynthesis \cite{arn07}. 
From the theoretical point of view, it
is desirable to have models which can describe these excitations 
in every nucleus, also in  
systems too short-lived to allow for experimental studies.

Collective nuclear excitations have been successfully described by the
Random Phase Approximation (RPA) theory \cite{spe91} whose extension
to handle charge-exchange excitations was formulated some time ago
\cite{hal67,lan80,aue81,aue83}. The description of available data has
been conducted by using a phenomenological input of the RPA, where
the single particle (s.p.) wave functions and energies were generated
by mean-field potentials, for example of Woods-Saxon type, and the
effective nucleon-nucleon interaction had Landau-Migdal form
\cite{spe91}. From these studies it has been possible to select the
value of the parameter defining the spin-isospin dependent term of the
interaction \cite{ost92,ich06}. The application of this
phenomenological approach is limited to nuclei whose ground state
properties are experimentally known. The theoretical exploration of
nuclei in the experimentally unknown regions of the nuclear chart
requires a more microscopic approach.

In this perspective, the combination of Hartree-Fock (HF) and RPA
calculations carried out with a unique effective interaction has been
able to provide a good description of known nuclear properties in a
wide range of the nuclear chart, from light nuclei, around the oxygen
region, up to very heavy nuclei such as the uranium. This success
induces to believe that this computational scheme can provide good
predictions of the properties of exotic nuclei which will be produced
in the next few years in radioactive ion beams facilities. This
possibility has increased the interest in defining more precisely the
details of the self-consistent HF+RPA calculations.

Self-consistent studies of charge-exchange excitations have been
conducted mainly with zero-range Skyrme-type interactions
\cite{ham00,fra07}. Recently, these interactions have been
implemented with tensor terms, and the effects of these new terms on
charge-exchange excitations have been studied
\cite{bai09a,bai09b,bai10,bai11a,bai11b,min13}. 
Other authors have studied charge-exchange 
excitations within quasi-particle RPA using the Bonn-A 
two-body potential in Woods-Saxon s.p. bases. \cite{bes12,civ14}.

In this work we apply a HF+RPA computational scheme based on
Gogny-like finite-range interactions to study isobaric analog states, 
Gamow-Teller, spin-quadrupole and spin-dipole excitations in \caIIp,
\zr and \pbp.  
Our first task is to test the validity of our model
against the available experimental data. We use four parametrizations
of the Gogny-like interaction. The D1S \cite{ber91} and D1M
\cite{gor09} forces contain the traditional set of parameters of the
original Gogny interaction.  Following the procedure outlined in
Ref. \cite{gra13} we add to these two parametrizations a tensor
force, and we call D1ST2c and D1MT2c these new interactions. The second
task of the present work is the study of the tensor effects on the
various observables related to charge-exchange excitations. 

The paper is structured as follows.  In Sec. \ref{sec:model} we
briefly present our HF+RPA model, and also the quantities we calculate
to compare our results with the experiment. The interactions we use in
this investigation are discussed in Sect. \ref{sec:details}. In the
same section we provide some information about the numerical details
of the calculations. We dedicate Sect. \ref{sec:results} to the
presentation of our results. In the first part of the section we
compare those results obtained by using interactions with and without
tensor forces. In the second part we study the effects of the various
terms of the interaction on the different charge-exchange excitations
we have considered. Finally, in Sect. \ref{sec:summary} we summarize
the main results of our investigation and we draw our conclusions.

\section{Model}
\label{sec:model}
The RPA theory describes the excited state of a many-body system as a
linear combination of particle-hole and hole-particle 
excitations. Because the states of the nucleus are
characterized by a total angular momentum $J$, it is convenient to work 
in an angular momentum coupling scheme,
where the RPA excited states are eigenstates of
the $\boldsymbol J ^2$ and $J_z$ operators:
\beq
 | JM \rangle 
 \, = \, \sum_{qk}\left[X_{qk}^J \, {\cal A}^{\dag}
 _{qk}(JM) \, 
-\, Y_{qk}^J \, \tilde{\cal A}_{qk}(J{M}) \,  \right]   |0 \rangle \, ,
\label{eq:nu}
\eeq
where $X^J$ and $Y^J$ are the RPA amplitudes and we have defined
\beqn
{\cal A}^{\dag}
_{qk}(JM)&=&\sum_{\mu_q \mu_k}\, \langle j_q \mu_q j_k \mu_k |JM \rangle \, a^{\dag }_{j_q, \mu_q}\, (-1)^{j_k+\mu_k} \, a_{j_k,-\mu_k} \, ,
\label{eq:xj}\\
\tilde{\cal A}_{qk}(J{M})&=&\sum_{\mu_q \mu_k} \, (-1)^{J-M}\, \langle j_q \mu_q j_k \mu_k|J-M\rangle  \,a^{\dag} _{j_k, \mu_k} \, (-1)^{j_q+\mu_q} \, a_{j_q,-\mu_q}
\, .
\label{eq:yj}
\eeqn
In the above equations $a^{\dag}$ and $a$ indicate the usual creation
and annihilation single nucleon operators, $k$ the quantum numbers
characterizing a s.p. state below the Fermi surface and $q$ those of a
state above it.  Besides, $j$ and $\mu$ are,
respectively, the angular momentum and its projection on the $z$-axis
of the nucleon. In the above 
expressions, we understood the explicit dependence 
of the excited state and of the $X$ and $Y$ amplitudes
on the parity $\Pi$ and on the excitation energy $\omega$.

Charge-exchange excitations can be classified as isospin lowering \tm
when the hole is a neutron and the particle is a proton, and isospin
rising \tp when the hole is a proton and the particle is a neutron.
We use the usual convention of $p$ and $n$ for a proton and a neutron
state, respectively, and the bar to indicate a hole state, therefore
we have $p \bn$ pairs in \tm, and  $n \bpp$ pairs in  \tp excitations.

Charge conservation allows to write the secular RPA equations in 
a compact form \cite{lan80,aue81}.  
We define two new variables $U^J$ and $W^J$  such as
in the \tm channel we have
\beq
X^J_{p\bn}\, =\, U^J_{p\bn}\, , \, \, \, Y^J_{n\bpp}\, =\, W^J_{n\bpp} \, \, \, {\rm
  and} \,\,\, \omega\,=\,\Omega \, , 
\eeq 
and in the \tp channel 
\beq
X^J_{n\bpp}\,=\,W^J_{n\bpp}\, , \, \, \, Y^J_{p\bn}\,=\,U^J_{p\bn} \, \, \, {\rm and}
\,\,\, \omega\,=\,-\Omega \, .  
\eeq
where we have indicated with $\omega$ the RPA excitation energy.
The normalization of the RPA excited states (\ref{eq:nu}) implies
\beq
\sum_{p\bn}\, (U^J_{p\bn})^2\,-\, \sum_{n\bpp} \,(W^J_{n\bpp})^2\, =\, \pm 1  \, ,
\eeq
where the plus sign is for the \tm excitations and the minus sign for \tp ones. 

With these definitions we write the RPA secular equations as
%
\beqn
\begin{bmatrix} A^J_{p\bn p' \bnI} &B^J_{p \bn n' \bpI} \\ 
-B^J_{n\bpp p' \bnI} & -A^J_{n \bpp n' \bpI} \\ 
\end{bmatrix}
\begin{bmatrix}
U^J_{p'\bnI}\\
W^J_{n'\bpI}\\
\end{bmatrix}=\Omega
\begin{bmatrix}
U^J_{p\bn}\\
W^J_{n\bpp}\\
\end{bmatrix} 
\label{eq:mat2}
\, .
\eeqn

where $A$ and $B$ are expressed in terms of the interaction matrix
elements and s.p. energies as: 
\beqn
\label{eq:AJ}
A^J_{abcd}&=&(\epsilon_a-\epsilon_b)\, \delta_{bc}\, \delta_{ad}\,+\, \overline{V}^J_{abcd}\, ,\\
\label{eq:BJ}
B^J_{abcd}&=&(-1)^{j_c-j_d-J} \, \overline{V}^J_{abdc}\, .
\eeqn
In the above equations, we have indicated with $\epsilon$ the
s.p. energies, and with the symbol $\overline{V}^J_{abcd}$ the
antisymmetrized matrix element of the interaction:
\beqn
\nonumber
\overline{V}^J_{abcd}&=& \sum_K \,(-1)^{j_b+j_{c}+K} \, \sqrt{2K+1}
\left\{\begin {array}{ccc}
j_a     &  j_b    &   J\\
j_c &  j_d &  K\\
\end{array}\right\}\\ && \hspace*{0.7cm}\left[
\langle j_a j_d K \| V \| j_b j_c K\,\rangle \,
-\, (-1)^{j_b+j_c+K}\,\langle
j_a j_d K \| V \| j_c j_b K\rangle 
\right]
\, .
\eeqn
In the above equation, the double bar symbol $\|$
indicates the reduced matrix element of the 
angular part.

The diagonalization of the system (\ref{eq:mat2}) produces at the same
time the solutions for \tm and \tp excitations.  
For a given excitation multipole, the
charge-exchange RPA solution provides the set of excitation energies,
and, for each excited state, the full set of RPA amplitudes $X^{J}$ and $Y^{J}$.

The strength function of the transition between the ground state
and an excited state $|J^\pi;\omega \rangle$ of a nucleus with $A$ nucleons
induced by a one-body transition operator of the type
\beq
\hfp_{J^\pi,M}^{\alpha\pm} \, = \, \sum_{i=1}^A \, {\eta}_{J^\pi,M}^{\alpha\pm}(i) \, ,
\label{eq:fgen}
\eeq
can be expressed as
\beqn 
\nonumber \Gamma_{J^\pi}^{\alpha\pm}(\omega)
&=& \sum_M |\langle J^\pi,M ; \omega| \hfp_{J^\pi,M}^{\alpha\pm}| 0\rangle|^2 \\ 
&=& \left|
\sum_{qk} 
\left(X^{J}_{qk}\, 
\langle q \| {\eta}_{J^\pi}^{\alpha\pm} \| k \rangle \,+
\, (-1)^{j_q-j_k+J+1} \, Y^{J}_{qk}\, \langle k \| 
{\eta}_{J^\pi}^{\alpha\pm} \| q \rangle \right) \right| ^2 \, ,
\label{eq:strength}
\eeqn
where $M$ is the $z$ axis projection of $J$. 
In the second line of Eq. (\ref{eq:strength}), 
we have applied the Wigner-Eckart theorem \cite{edm57},
therefore we dropped the explicit dependence on $M$.
Also in this case, as in Eq.(\ref{eq:nu}), we understand 
the dependence of the $X$ and $Y$ amplitudes 
on the parity $\Pi$ and of the excitation energy $\omega$
 of the excitation.
Since we consider even-even nuclei only,
the angular momentum and the parity, $J^\pi$, of the
excitation coincide with those of the nuclear final 
state. 
We list here below the transition operators which we consider in this work. 
For the excitation of the $0^+$ states, the isobaric analog states, we consider the
Fermi (F) operator  
\beq
\hfp_{0^+,0}^{{\rm F}\pm}\, = \, \sum_{i=1}^A \,t_{\pm}(i)
\label{eq:f}
\, .
\eeq
For the excitation of the $1^+$ states we use the Gamow-Teller (GT) 
operator 
\beq
\hfp_{1^+,M}^{{\rm GT}\pm} \,= \, \sum_{i=1}^A  \, \bsigma_M(i) \, t_{\pm}(i)\, = \, 
\sqrt{4\pi} \, \sum_{i=1}^A  \, [Y_0(i) \otimes \bsigma(i) ]^1_M \, t_{\pm}(i)
\, ,
\label{eq:gt}
\eeq
and the spin quadrupole (SQ) operator
\beq
\hfp_{1^+,M}^{{\rm SQ}\pm} \, = \,\sum_{i=1}^A  \,r^2_i \,  [Y_2(i) \otimes \bsigma(i)  ]^1_M \,
t_{\pm}(i) 
\label{eq:sq}
\, .
\eeq
Finally, we consider the excitations induced by the 
spin dipole (SD) operator 
\beq
\hfp_{J^-,M}^{{\rm SD}\pm} \, = \, \sum_{i=1}^A \, r_i \, [Y_1(i)\otimes \bsigma(i) ]^J_M  \, 
t_{\pm}(i) 
\label{eq:sd}
\, ,
\eeq
which excites the multipoles $0^-$, $1^-$ and $2^-$. 
In this case, apart from the strength functions corresponding to each
individual multipolarity, also the total strength 
\beq
\Gamma^{{\rm SD}\pm}(\omega) \, = \, \sum_{J^\pi=0^-,1^-,2^-}
\Gamma_{J^\pi}^{{\rm SD}\pm}(\omega) 
\label{eq:sumSD}
\eeq
has been calculated. 
In the previous equations we used $t_\pm = \tau_\pm /2$ 
where $\tup$ and $\tdo$ are the
isospin operators transforming, in our convention, a
proton into a neutron and vice-versa, respectively.
Furthermore, we have indicated with
$Y_{L}$ the spherical harmonics and with
$\bsigma$ the Pauli matrix operator acting on the spin
variable.
The symbol $[A \otimes B]$ indicates the usual tensor
product between irreducible sperical tensors \cite{edm57}. 
The expressions of the reduced matrix elements of Eq. 
(\ref{eq:strength}) are given in Appendix \ref{sec:appb} 
for the operators we have presented above. 

The sum rules are an important tool to investigate the global
properties of the charge-exchange excitations. In order to obtain the
sum rule expressions, it is useful to define the energy moments:
\beq
m_{\lambda}^{\alpha\pm}\, =\, \sum_{J^\pi} \, m_{\lambda}(\hfp^{\alpha\pm}_{J^\pi}) \, ,
\label{eq:mom}
\eeq
where
\beq
m_{\lambda}(\hfp^{\alpha\pm}_{J^\pi})\, =\, \int_0^\infty {\rm d}\omega \, \omega^\lambda
\, \Gamma_{J^\pi}^{\alpha\pm}(\omega) 
\, .
\label{eq:momJ}
\eeq

According to these expressions,
we define the centroid energy of an excitation induced by 
the $\alpha$-type operator as
\beq
{\omega}_{\rm cen}^{\alpha\pm} \, =\, \frac{m_{1}^{\alpha\pm}}{m_{0}^{\alpha\pm}} \,   . 
\label{eq:centr}
\eeq
In the case of the SD transitions, 
we have also calculated the centroid of the distributions of the
individual multipolarities 
\beq
{\omega}_{{\rm cen},J^\pi}^{{\rm SD}\pm} \, 
=\, \frac{m_{1}(\hfp^{{\rm SD}\pm}_{J^\pi})}{m_{0}(\hfp^{{\rm SD}\pm}_{J^\pi})} \,   . 
\label{eq:centr-mul}
\eeq

By using the property $(\tau_\pm)^{\dag} = \tau_\mp$, 
and the completeness of the
RPA excited states we have that \cite{aue81,aue83}
\beqn
\nonumber
m_0^{\alpha-}\, -\, m_0^{\alpha+} &=& \sum_{J^\pi} \, \int_0^\infty {\rm d}\omega \,
\left( |\langle J^\pi;\omega|\hfp^{\alpha-}_{J^\pi}|0 \rangle|^2 \,
- \, |\langle J^\pi;\omega | \hfp^{\alpha+}_{J^\pi}|0 \rangle|^2 \right) \\
&=&\langle 0|[\hfp^{\alpha-}_{J^\pi},\hfp^{\alpha+}_{J^\pi}]|0 \rangle \, ,
\eeqn
which depends only on the nuclear ground state. 
In particular, the F operator satisfies the IAS sum rule \cite{ost92}
\beq
\dfer \equiv
m_0^{{\rm F}-} \,-\, m_0^{{\rm F}+}\,=\, N-Z 
\, ,
\label{eq:srF}
\eeq
which is the difference between neutron and proton numbers.
For the GT operator we have the well known sum rule, often
called Ikeda sum rule \cite{gaa80}, 
\beq
\dgt \equiv
m_0^{{\rm GT}-} \,-\, m_0^{{\rm GT}+}\,=\,3\, (N-Z) 
\, .
\label{eq:srGT}
\eeq
The SD transitions satisfy:
\beq
\dms \equiv
m_0^{{\rm SD}-}\, -\, m_0^{{\rm SD}+}\,
=\, \frac{9}{4\pi}\, \left[ N\, r^2_n \,-\, Z \, r^2_p \right]
\, ,
\label{eq:srSD}
\eeq 
where $r_n$ and $r_p$ are the mean square radii of
neutrons and protons, respectively.

\section{Details of the calculations}
\label{sec:details}

The only input required by our self-consistent approach
is the effective nucleon-nucleon force.
In this work we use Gogny-like interactions which are
composed by five
finite-range terms: the scalar, isospin, spin, spin-isospin
and  Coulomb terms.
These interactions contain, in addition, 
a density dependent and a spin-orbit zero-range terms. 
We carried out calculations with the D1M force \cite{gor09}, with the more
traditional D1S \cite{ber91} parametrization, and also with 
other two forces, which we built by adding tensor terms to the 
two basic parameterizations. 
In these new forces, which we name D1MT2c and D1ST2c, we did not
change any value of the parameters of the original D1S and D1M
interactions, but that related to the spin-orbit force.  
Following the
work of Refs. \cite{gra13,ang12,oni78}, we include two tensor
terms of the form
\beq
V_{\rm tensor}(i,j) \,=\, (V_{\rm T1} \,+\, V_{\rm T2}\, P^\tau_{ij}) \, S_{ij} \,
\exp \left[\displaystyle -\frac{(r_i-r_j)^2}{\mu^2_{\rm T}} \right]  \, ,
\label{eq:tensor}
\eeq
where $\mu_{\rm T}=1.2$ fm corresponds to the longest range used in
the D1M and D1S forces, 
$P^\tau_{ij}$ is the usual isospin exchange operator defined as
\beq
P^\tau_{ij} \, = \, \frac{1\, + \, \btau(i)\cdot\btau(j)}{2}
\,\,,
\label{eq:pt}
\eeq
and $V_{\rm T1}$ and $V_{\rm T2}$ are two constants. 
Eq.(\ref{eq:tensor}) can be rewritten as 
\beq
V_{\rm tensor}(i,j) \,=\, \left[V_{\rm T} \,+\, V_{{\rm T}\tau} \, \btau(i)\cdot\btau(j) \right] 
\, S_{ij} \,
\exp \left[\displaystyle -\frac{(r_i-r_j)^2}{\mu^2_{\rm T}} \right]
\, ,
\label{eq:tensor2}
\eeq
and, in the following, we shall call {\sl pure tensor} the term dependent on 
$V_{\rm T} = V_{\rm T1} + V_{\rm T2}/2$ 
and {\sl tensor-isospin} that dependent on $V_{{\rm T}\tau} = V_{\rm T2}/2$.
In the previous equations we have used the following definition of 
the tensor operator 
\beq
S_{ij}\, = \,3 \, \frac {\bsigma(i)\cdot \br_{ij} \,
                    \bsigma(j)\cdot \br_{ij} } 
                 {r_{ij}^2}\,
- \, \bsigma(i)\cdot\bsigma(j) \, ,
\eeq
where
\beq
\br_{ij} \, = \, \br_i \, - \, \br_j 
\label{eq:r}
\eeq 
represents the relative coordinate of the two interacting nucleons. 

We select the values of  $V_{\rm T1}$ and $V_{\rm T2}$ by following the procedure
described in Ref. \cite{gra13} consisting in reproducing the
experimental energy splitting between the neutrons $1f_{7/2}$ and $1f_{5/2}$
states in $^{40}$Ca, \caII and $^{56}$Ni nuclei, whose values are 6.8, 8.8
and 7.16 MeV, respectively \cite{firestone}. These observables are also
sensitive to the spin-orbit term of the force, whose strength is
characterized by the parameter $W_{\rm LS}$. The number of
experimental data we have reproduced corresponds to the
number of the free parameters we have to choose.  The values of the tensor
and spin-orbit parameters which characterize the D1MT2c and D1ST2c
forces are given in Table \ref{tab:tensor}. 

\begin{table}[t]
\begin{center}
\begin{tabular}{cccc}
\hline \hline 
      & $V_{\rm T1}$ [MeV]&  $V_{\rm T2}$ [MeV] & $W_{\rm LS}$  [MeV fm$^5$] \\
\hline
D1ST2c  & -135 & 60 &  103 \\
D1MT2c  & -175 & 40 & 95\\
\hline\hline
\end{tabular}
\caption{Values of the parameters of the tensor force, 
given in  Eq. (\ref{eq:tensor}), and of the spin-orbit term of the
nucleon-nucleon 
interactions considered in the present work.
}
\label{tab:tensor}
\end{center}
\end{table}

\begin{figure}[h]
\begin{center}
\parbox[c]{16cm}{\includegraphics[scale=0.4,angle=0.0]{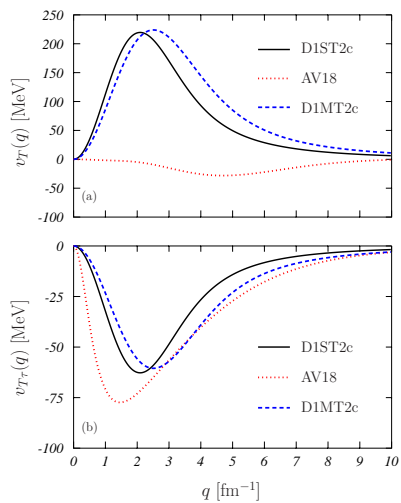}}
\caption{\small (Color online) 
Pure tensor (panel (a)) and tensor--isospin (panel (b)) 
terms of the D1ST2c and
D1MT2c parameterizations used in this work
as a function of the relative momentum of the interacting nucleon
  pair, compared with the analogous terms of the realistic 
 interaction AV18 \cite{wir95}.}
\label{fig:compare}
\end{center}
\end{figure} 

We show, in Fig. \ref{fig:compare}, the pure tensor, $v_{\rm T}$,
  and tensor-isospin, $v_{{\rm T}\tau}$, terms of the D1ST2c and
  D1MT2c forces as a function of the relative momentum of the two
  interacting nucleons, and we compare them with the analogous terms
  of the microscopic Argonne V18 (AV18) interaction \cite{wir95}. In
  this figure, the differences between the microscopic and, our,
  effective interactions become evident. The AV18 interaction has an
  actractive tensor-isospin term almost three times larger than the
  pure tensor term, which is again attractive. If the tensor-isospin
  terms of our effective interaction are similar to that of the AV18, 
  the pure tensor terms are remarkably different. The sign is
  different, these terms are repulsive instead than actractive. In
  addition their size is much larger than that of the analogous AV18
  term  and also, in absolute value, of that of the tensor-isospin
  terms. Understanding the origin of these differences is an
  interesting topic, but we do not tackle it in this paper. We take
  for granted our effective interaction and we are interested in
  identifying eventual observables in charge-exchange excitations
  which are sensitive to the presence of the tensor force.

The first step of our calculations consists in constructing the
s.p. basis by solving the HF equations with the bound-state boundary
conditions at the edge of the discretization box.  The technical
details concerning the iterative procedure used to solve the HF
equations for a density-dependent finite-range interaction can be
found in Refs. \cite{co98b,bau99}. 
When the stable solution,
corresponding to the minimum of the binding energy, is reached,
we construct the local Hartree and the non local Fock-Dirac
potentials by using the s.p. wave functions lying below the Fermi
surface. By using these potentials, we solve the HF equations also
for those states above the Fermi surface.
In this way, we generate a set of discrete bound states also
in the positive energy region, which should be characterized by the
continuum. The level density in the continuum region is strictly
related to the size of the space integration box: the larger is the
box the higher is the level density.

The second step of our calculations consists in solving the RPA secular
equations by diagonalization.  The explicit expressions of the $A^J$
and $B^J$ matrix elements in Eqs. (\ref{eq:AJ}) and (\ref{eq:BJ}) for
Gogny interactions can be found in Refs.  \cite{don14,don09}.  The
dimensions of the matrix to diagonalize are given by the sum of the $p
\bn$ and $n \bpp$ pairs that depends on the number of the s.p. states
composing the configuration space.

In our approach, the stability of the RPA results depends on two parameters:
the size of the integration box, and the maximum s.p. energy.  We have
chosen the values of these two parameters by controlling that the
centroid energies of the electric giant dipole 
resonances in charge conserving
RPA do not change by more than 0.5 MeV when either the box size or the
maximum s.p. energies are increased.  We have done calculations for
the \caIIp, \zr and \pb nuclei. The most demanding calculations are
those we carried out for the \pb nucleus.  In this nucleus, by using a
box radius of 25 fm and an upper limit of s.p. energy of 100 MeV, we 
diagonalized matrices of dimensions of about 1300 $\times$ 1300.

Our HF+RPA calculations are fully self-consistent, we have
used for the evaluation of the RPA excited states the same interaction
adopted to generate the s.p. wave functions and energies, including the
Coulomb and the spin-orbit channels.  Of course, the former
interaction is not active in charge-exchange excitations. These terms
of the effective nucleon-nucleon interaction are usually neglected in
RPA calculations, since the evaluation of their contribution,
considered small as compared to that of the other terms of the
interaction, is computationally quite heavy. 
Recently, we studied the relevance of these two terms of the interaction
in charge conserving HF+RPA calculations \cite{don14}.

\begin{table}[htb]
\begin{center}
\begin{tabular}{crcccccl}
\hline \hline 
 & & D1M & D1MT2c & D1S & D1ST2c &~~& exp\\
\hline
{\caII} & $B/A$ & 8.590 & 8.614 & 8.690 & 8.632 && 8.667 \cite{aud03,audw}\\
\cline{2-8}
&$r_n $&3.550 & 3.552 & 3.586 & 3.597 & &--\\
&$r_p$ &3.415 & 3.418 & 3.441 & 3.460 & &-- \\
&$r_c$ &3.525 & 3.528 & 3.548 & 3.557 & &3.451 $\pm$ 0.009 \cite{vri87}\\
\cline{2-8}
&$\drnp$&  0.135 &  0.134 & 0.145  & 0.097& &--\\
\hline
\zr & $B/A$ & 8.636 & 8.670 & 8.736 & 8.692 & &8.710 \cite{aud03,audw}\\
\cline{2-8}
&$r_n$& 4.231   & 4.230 & 4.269  & 4.277   & &--\\
&$r_p$& 4.179   & 4.177 & 4.209  & 4.217   & &--\\
&$r_c$& 4.269   & 4.269 & 4.298  & 4.305   &  &4.258$\pm$ 0.008 \cite{vri87}\\
\cline{2-8}
&$\drnp$ & 0.052  & 0.053 &0.060 & 0.060 & &0.09 $\pm$ 0.07 \cite{ray78} \\
  &    &       &      &      &       & &0.07 $\pm$ 0.04 \cite{yak06}  \\
\hline
\pb & $B/A$ & 7.830 & 7.815 & 7.889 & 7.801 & &7.867 \cite{aud03,audw}\\
\cline{2-8}
&$r_n$ & 5.505  & 5.514 & 5.554 & 5.570 &&$5.78^{+0.16}_{-0.18}$ \cite{abr12} \\
&$r_p$ & 5.413  & 5.420 & 5.433 & 5.446 && --\\
&$r_c$ & 5.480  & 5.488 & 5.498 & 5.512 & &5.503$\pm$ 0.002 \cite{vri87}\\
\cline{2-8}
&$\drnp$   &0.092   &  0.094 & 0.121 &  0.124  & &$0.083<\drnp<0.111$
\cite{cla03} \\
&        &       &      &        &         & &0.19 $\pm$ 0.09 \cite{kra91}\\
\hline \hline 
\end{tabular}
\caption{
Binding energies per nucleon, $B/A$, in MeV, and
neutron, $r_n$, proton, $r_p$, and charge, $r_c$, rms radii, in fm, 
of $^{48}$Ca, \zr and \pb nuclei obtained in HF calculations by using
the various interactions adopted in our work. 
}
\label{tab:radii}
\end{center}
\end{table}

\section{Results}
\label{sec:results}
In this section we show some results of our investigation of 
charge-exchange excitations of three nuclei: \caIIp, \zr and \pbp.  We
present results related to F, GT, SQ and SD excitations.
First, we address our attention to the
differences between the results obtained with and without 
the tensor force. Since the tensor effects are rather similar for the
two types of forces considered, we show
in the figures only the strength distributions
obtained by using the D1M and D1MT2c interactions. 
In the tables we present
global results of our calculations obtained with all the interactions
considered.

As pointed out in the previous section, the first step of our approach
is the generation of the s.p. configuration space for each nucleus
considered by means of a HF calculation. 
We present in Table \ref{tab:radii}
some results of these calculations: the binding energies per nucleon, $B/A$,
the neutron, $r_n$, proton, $r_p$, charge $r_c$, root mean square (rms) radii, 
and the neutron skin, $r_n-r_p$.  
The charge distributions used to extract the
$r_c$ radii, have been obtained by folding the pointlike proton
distributions with a dipole proton electromagnetic form factor. The
use of more refined form factors changes the radius values of few
parts on a thousand. The experimental values of the binding energies
have been taken from Refs. \cite{aud03,audw} and those of the charge
radii from the compilation of Ref. \cite{vri87}. The empirical value
of the \pb neutron rms radius has been obtained by the parity
violation electron scattering PREX experiment \cite{abr12} and those
of the neutron skins from Refs. \cite{ray78,yak06} for \zrp, and
\cite{cla03,kra91} for \pbp.

The agreement with the available experimental data is, in general,
very good. This is not surprising since the binding energies and rms
radii are part of the set of data used in the fit procedure adopted to
select the values of the parameters of the D1M and D1S forces
\cite{cha07t}. We observe that the inclusion of the tensor forces does
not modify sensitively the values of these observables.

\begin{table}[b]
\begin{center}
\begin{tabular}{cccccccc}
\hline \hline
  &  &  & D1M & D1MT2c & D1S & D1ST2c & exp\\
\hline
{\caII} &$p$ & $1f_{7/2}$ & -9.83 & -8.44 & -9.90 & -8.18 & -9.45\\
&$\bn$ & $1f_{7/2}$ & -9.33 & -9.72 & -9.48 & -9.68 & -9.94\\
\hline
{\zr}&$p$  &$1g_{9/2}$ & -5.78 & -4.45 & -5.98 & -4.36 & -5.08\\
&$\bn$  &$1g_{9/2}$ & -11.80 & -12.10 & -11.90 & -12.02 & -11.97\\
\hline
{\pb}&$p$  &$1h_{9/2}$ & -3.33 & -3.87 & -3.56 & -4.21 & -3.71\\
&$\bn$  &$3p_{1/2}$ & -8.94 & -8.29 & -7.85 & -8.09 & -7.37\\
\hline\hline
\end{tabular}
\caption{Energies, in MeV, of the s.p. states near the Fermi surfaces
of \caII and \zr and \pb nuclei. We present the energies of the first   
proton empty state ($p$), and that of the last
neutron occupied state ($\bn$). The experimental
values have been taken from the compilation of Ref. \cite{firestone}.
}
\label{tab:spene}
\end{center}
\end{table}

In Table \ref{tab:spene}, for each of the three nuclei considered,
we show the s.p. energies of the last
occupied neutron  states and the first empty proton
states. It is evident that the HF calculations done with D1M and D1S
interactions generate \caII ground states which are unstable under
beta decay, since the energies of the unoccupied proton $1f_{7/2}$
state are lower than those of the analogous, occupied, neutron state.
This instability of the HF ground state against the beta decay is not
present in the other nuclei.
The inclusion of tensor terms solves this problem, as it is shown by
the s.p. energies corresponding to the D1ST2c and D1MT2c forces given
in the table.  We point out that the parameters of these interactions
have been chosen to reproduce other observables, {\it i.e.} the spin-orbit
splitting of the $1f$ neutron states in \caI, \caII and $^{56}$Ni,
therefore this is a genuine prediction of our model.

\begin{table}[htb]
\begin{center}
\begin{tabular}{cccccccl}
\hline
\hline
 &  & D1M & D1MT2c & D1S & D1ST2c &~~&{expected}\\
\hline
\caII & $m_0^{\rm F-}$ &  8.49  & 8.34 & 8.40 & 8.26 & &\\
        & $m_0^{\rm F+}$ & 0.49  & 0.34 & 0.40 & 0.26 & &\\
        & $\dias$               & 8.00  & 8.00 & 8.00 & 8.00 & & 8.00\\ 
\hline
\zr    & $m_0^{\rm F-}$ &  10.59  & 10.41 & 10.49 & 10.33 &&\\
        & $m_0^{\rm F+}$ & 0.60  & 0.41 & 0.49 & 0.33 &&\\
        & $\dias$               & 9.99  & 10.00 & 10.00 & 10.00 && 10.00\\ 
\hline
\pb   & $m_0^{\rm F-}$ &  46.40  & 46.66 & 46.33 & 46.40 &&\\
        & $m_0^{\rm F+}$ & 2.60  & 2.66 & 2.33 & 2.40 &&\\
        & $\dias$               & 43.80  & 44.00 & 44.00 & 44.00 && 44.00\\ 
\hline
\hline
 \end{tabular}
\caption{The IAS sum rule values for F excitations in
  \caIIp, \zr and \pb nuclei. The RPA responses 
  have been integrated up to the maximum energy of 250 
  MeV for $^{48}$Ca, 200 MeV for $^{90}$Zr and 150 MeV for $^{208}$Pb.
    The $\dias$ values in the columns
  labelled with the force name are obtained as the difference
    between $m_0^{\rm F-}$ and $m_0^{\rm F+}$, while
  the expected values are $N-Z$ (see Eq. (\ref{eq:srF})).
}

\label{tab:IASsr}
\end{center}
\end{table}

We start our discussion about the charge-exchange excitations by
considering first the IAS resonance.  The validity and the consistency
of our RPA calculations can be verified by observing the exhaustion of
the sum rules (\ref{eq:srF}) whose values are given in
Table \ref{tab:IASsr}.  
The good agreement with the expected values
indicates that our configuration spaces are large enough
to reach the numerical convergence of our calculations.  As expected,
in nuclei with neutron excess,  
the total strength carried by the \tm excitation is much larger than
that of the \tp excitation.

The IAS resonances in the \caII and \zr nuclei are dominated by the
neutron-proton transitions between the analog $1f_{7/2}$ states in
\caIIp, and $1g_{9/2}$ states in \zrp.  In RPA calculations the IAS
excitation presents a well isolated large peak which carries more than
the 90\% of the total strength.  Also in \pb the IAS strength
distribution shows a single sharp peak, however the situation is more
complicated since there are various particle-hole (p-h) excitations
contributing to the main excitation. The energies of the IAS peak,
$\omega^{\rm F-}_{\rm max}$, for each nucleus and interaction
considered are compared in Table \ref{tab:centroid} with the
experimental values extracted from Refs.
\cite{bai80,and85,wak97,aki95,wak05,yak06,yak09,wak10,wak12}.

\begin{table}[b]
\begin{center}
\begin{tabular}{cccccccl}
\hline
\hline
 &  & D1M & D1MT2c & D1S & D1ST2c &~~& exp\\
\hline
\caII & $\omega^{\rm F-}_{\rm max}$ & 5.67 & 6.26 &5.66 & 6.25 && $7.17$ \cite{and85}\\
 && {\it -0.50}& {\it 1.28} & {\it -0.52} & {\it 1.50} & & {\it (IPM)}  \\\cline{2-8}
&$\omega^{\rm GT-}_{\rm max}$ & 11.64& 9.90 & 12.38 & 10.17 && 10.5 \cite{and85}\\
&$\omega^{\rm GT-}_{\rm cen}$ &9.87 & 10.35 & 10.28 & 10.26 && --\\\cline{2-8}
&$\omega^{\rm SD-}_{\rm cen}$ &21.61 & 22.48 & 20.81 &  20.83 && -- \\
\hline
\zr &$\omega^{\rm F-}_{\rm max}$ & 10.88 & 11.32 & 10.79 &  11.19 & &$12.0 \pm 0.2$ \cite{bai80} \\
 && {\it 6.02}& {\it 7.65} & {\it 5.92} & {\it 7.66} & & {\it (IPM)}  \\\cline{2-8}
&$\omega^{\rm GT-}_{\rm max}$ & 17.36 & 15.64 & 17.93 & 15.80& &15.6 $\pm$ 0.3 \cite{wak97,wak05}\\
&$\omega^{\rm GT-}_{\rm cen}$ &15.68 &15.80 & 15.84 &15.90 & &16.54 \cite{wak05} \\\cline{2-8}
&$\omega^{\rm SD-}_{\rm cen}$ &24.98 & 24.47 & 25.56 &  24.90 && 30.74 \cite{yak06} \\
\hline
\pb &$\omega^{\rm F-}_{\rm max}$ & 17.23& 17.21 & 17.02 & 16.97 & &$18.83\pm0.02$ \cite{aki95} \\
 & & {\it 11.35}& {\it 11.32} & {\it 10.77} & {\it 10.74} & & {\it (IPM)}\\\cline{2-8}
&$\omega^{\rm GT-}_{\rm max}$ & 20.99 & 18.89 & 21.12 & 18.56 && $19.2 \pm 0.2$ \cite{aki95} \\
&$\omega^{\rm GT-}_{\rm cen}$ &19.64 & 18.74 & 19.02 & 19.61 & & -- \\\cline{2-8}
&$\omega^{\rm SD-}_{\rm cen}$&25.08 & 25.04 & 25.40 &  25.01&&  28.37 \cite{wak10,wak12} \\
\hline\hline
\end{tabular}
\caption{Main peak, $\omega^{\alpha-}_{\rm max}$, and  
centroid, ${\omega}_{\rm cen}^{\alpha-}$, energies, in MeV, for the F, 
GT, and SD responses.
In the case of the F transitions, also the IPM 
values of the peak energies are shown, in italic. 
}
\label{tab:centroid}
\end{center}
\end{table}

Only the RPA calculations can provide a realistic description of these
IAS excitations.  The excitation energies in a pure independent
particle model (IPM) can be obtained as the difference between the
energies of the neutron and proton analog s.p. states.  For \caII and
\zrp, these energies are those shown in Table \ref{tab:spene}.  In the
case of \pbp, we considered the $\bar{n}(3p_{1/2})$ and $p(3p_{1/2})$
s.p. states.  The IPM values obtained are shown for the three nuclei
in Table \ref{tab:centroid} (in italic).  We observe indeed that the
IAS energies obtained in the IPM are extremely small with respect to
those of the RPA, and they are even negative for \caII in the D1M and
D1S cases, as we have pointed out above.  This indicates that
interactions and RPA correlations play an important role in the
description of these excitations, even though they are not collective
states, indeed their strength is concentrated in a single resonance
largely dominated by the IPM p-h transition.

In \caII and \zr the inclusion of the tensor force increases the
values of peak energy $\omega^{\rm F-}_{\rm max}$ in the correct
direction to improve the description of the experimental value by
about 0.5 and 0.4 MeV respectively.  The effect of the tensor force is
smaller and of opposite sign in the \pb nucleus. 
Though the quality of the description of
the experimental peak energies is not satisfactory, it is, however,
similar to that obtained by self-consistent calculations carried out
with Skyrme interactions \cite{col94}.

\begin{figure}
\begin{center}
\includegraphics[scale=0.4] {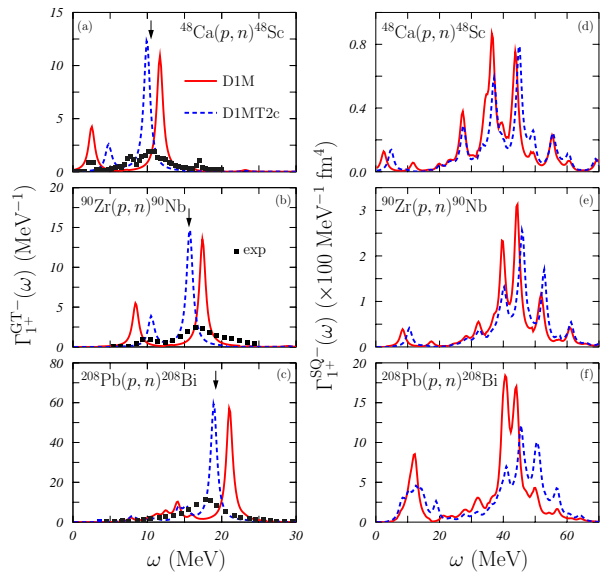}
\caption{\small  (color on line) 
Energy distributions of the
$\Gamma_{1^+}^{{\rm GT}-}(\omega)$ 
(left panels), and $\Gamma_{1^+}^{{\rm SQ}-}(\omega)$ (right panels) 
strengths, as given by Eq. (\ref{eq:strength}). 
The red solid curves have been obtained with the D1M interaction 
while the blue dashed curves with the D1MT2c force. 
The arrows indicate the experimental energies of the main peaks 
\cite{and85,wak97,aki95,wak05,wak10,wak12}. 
The black squares show the 
$^{48}$Ca, $^{90}$Zr and $^{208}$Pb experimental 
data, taken, respectively, 
from Refs. \cite{yak09,wak05,wak12}. }
\label{fig:gt}
\end{center}
\end{figure}

The role of the tensor force is more relevant in the GT excitations.
We give in Table \ref{tab:centroid} the energies of the main peaks,
$\omega^{\rm GT-}_{\rm max}$, and the centroid energies $\omega^{\rm
  GT-}_{\rm cen}$, of this type of excitation for the three nuclei and
for all the interactions considered.  In the left panels of
Fig. \ref{fig:gt} we present the $\Gamma_{1^+}^{{\rm GT}-}(\omega)$
strength distributions obtained with the D1M (red solid curves) and
D1MT2c (blue dashed curves) interactions.  In the figure, our discrete
results have been folded with a Lorentz function of 1 MeV width. The
arrows indicate the experimental values of the main peak energies
\cite{and85,wak97,aki95,wak05}. The consistency and convergence of our
calculations can be verified by observing the sum rule values given in
Table \ref{tab:GTsr}. The results shown in this table indicate that
the \tm transitions carry the major part of the total sum rule.

\begin{table}[b]
\begin{center}
\begin{tabular}{cccccccl}
\hline
\hline
 &  & D1M & D1MT2c & D1S & D1ST2c &~~&{expected}\\
\hline
\caII & $m_0^{\rm GT-}$ &  24.74  & 24.64 & 24.57 & 24.47 & &\\
        & $m_0^{\rm GT+}$ & 0.74  & 0.71 & 0.57 & 0.64 & &\\
        & $\dgt$               & 24.00  & 23.93 & 24.00 & 23.83 & & 24.00\\ 
\hline
\zr    & $m_0^{\rm GT-}$ &  31.10  & 30.92 & 30.92 & 30.72 &&\\
        & $m_0^{\rm GT+}$ & 1.12  & 0.99 & 0.92 & 0.82 &&\\
        & $\dgt$               & 29.98  & 29.93 & 30.00 & 29.90 && 30.00\\ 
\hline
\pb   & $m_0^{\rm GT-}$ &  137.33  & 135.97 & 136.91 & 134.92 &&\\
        & $m_0^{\rm GT+}$ & 5.38  & 5.26 & 4.96 & 4.90 &&\\
        & $\dgt$               & 131.95  & 130.71 & 131.95 & 130.02 && 132.00\\ 
\hline
\hline
 \end{tabular}
\caption{Sum rule for GT excitations in \caIIp,
  \zr and \pb nuclei.  The $\dgt$ values in the columns
  labelled with the force name are obtained as 
 the difference  between  $m_0^{\rm GT-}$ and $m_0^{\rm GT+}$, while
  the expected values are $3(N-Z)$ (see Eq. (\ref{eq:srGT})).
  The values of the maximum excitation energies 
  used to integrate the RPA responses are the same as those 
  indicated in the caption of table \ref{tab:IASsr}.}
\label{tab:GTsr}
\end{center}
\end{table}

From Fig. \ref{fig:gt} we observe that all the GT strength
distributions present essentially two peaks.  The smaller ones lie
well below the experimental energy of the main peaks.  The largest
peaks obtained with the D1M interaction occur at energies close,
slightly above, these experimental values.  The use of the D1MT2c
force, which includes the tensor terms, changes the position of these
peaks, even though does not modify sensitively the values of the
centroid energies, as it is shown in table \ref{tab:centroid}.  The
tensor terms reduce the energy of the large peaks by 2-3 MeV and
remarkably improve the agreement with the experimental data. These
effects of the tensor force are similar to those found with Skyrme
interactions \cite{bai11a}.  As seen in the panel (b) of the figure,
our RPA results describe reasonably well the positions of the peaks
but they miss completely the description of the experimental energy
distribution of the strengths.  This is not a specific problem of our
implementation of the HF+RPA approach, but rather an intrinsic limit
of the RPA that, by considering 1p-1h excitation only, does not
include the spreading width.  The experimental data for $^{90}$Zr
  may contain the contribution of the excitation induced by the
  isovector spin monopole operator \cite{boh75,conde92}. The discussion
  done in Ref. \cite{wak97} indicates that the presence of this type
  of excitation is negligible in the data measured in the experiment at
  forward scattering angle, and for this reason we did not consider
  it.  It is however a topic worth to be further investigated
  \cite{bes12,civ14}. 

Using Skyrme interactions, Bai {\it et al.}  \cite{bai09} showed that
about 10\% of the GT strength is moved above 30 MeV when the tensor
terms are included in the RPA calculation.  We have analyzed the
strength distributions of our GT results and have found a similar
effect though the shift of the strength is only 5\%.

To make another comparison with the results of Ref. \cite{bai09} we
have calculated the $\Gamma_{1^+}^{{\rm SQ}-}(\omega)$ strength
distributions obtained with the SQ operator. In the right panels of
Fig. \ref{fig:gt} we show the results obtained with the D1M (red solid
curves) and D1MT2c (blue dashed curves) forces. The
basic effect of the tensor force is to move the strengths towards
higher energies.
The sizes of these shifts  are much smaller than those found
in Ref. \cite{bai09,sag14} for the Skyrme interaction.

\begin{figure}[t]
\begin{center}
\includegraphics[scale=0.4] {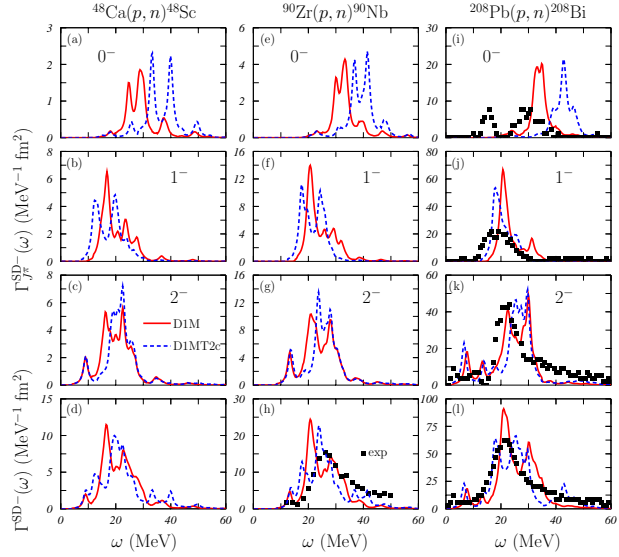}
\caption{\small  (color on line) 
Energy distributions of the
$\Gamma_{J^\pi}^{{\rm SD}-}(\omega)$ and 
$\Gamma^{{\rm SD}-}(\omega)$ strengths for \caII (left panels), 
\zr (central panels) and \pb (right panels) nuclei.  
The red solid curves have been obtained with the D1M interaction 
while the blue dashed curves with the
D1MT2c force. The 
squares indicate the experimental data of Ref. \cite{yak06}, for $^{90}$Zr, 
and of Refs. \cite{wak10,wak12}, for \pbp.
   }
\label{fig:sd}
\end{center}
\end{figure}

In Fig. \ref{fig:sd} we show the $\Gamma_{J^\pi}^{{\rm SD}-}(\omega)$
strength distributions for \caII (left panels), \zr (central panels)
and \pb (right panels) nuclei.  The results obtained with the D1M (red
solid curves) and the D1MT2c (blue dashed curves) interactions are
shown.  These excitations imply the superposition of the responses of
three different multipoles, the $0^-$ (panels (a), (e) and (i)), $1^-$
(panels (b), (f) and (j)), and $2^-$ (panels (c), (g) and (k)). Our
calculations produce discrete results for each multipole considered
even above the nucleon emission threshold. Since the experimental
strengths of Refs. \cite{yak06,wak10,wak12} are above this threshold,
a comparison with them requires the sum of the three responses.  
Because of the large numbers of peaks in this excitation region 
we fold our discrete results with a Lorentz function of 2MeV width. 
This procedure produces smooth continuous strength distributions which 
we sum for each value of the excitation energy, as indicated in Eq. (\ref{eq:sumSD}),
to obtain the total strength $\Gamma^{{\rm SD}-}(\omega)$ (panels
(d), (h) and (l)).

\begin{table}[htb]
\begin{center}
\begin{tabular}{cccccccl}
\hline
\hline
 &  & D1M & D1MT2c & D1S & D1ST2c & ~~&{exp}\\
\hline
\caII & $m_0^{\rm SD-}$ &  137.60  & 141.52 & 140.08 & 145.68 & &--\\
        & $m_0^{\rm SD+}$ & 51.96  & 53.64 & 51.81 & 56.39 & &--\\
        & $\dms$               & 85.64  & 87.88 & 88.27 & 89.29 & & --\\ 
expected & $\dms$ & 85.68  & 85.67 & 88.27 & 88.89 & & --\\ 
\hline
\zr    & $m_0^{\rm SD-}$ &  276.42  & 279.53 & 281.51 & 285.41 && $271\pm 14$\\
        & $m_0^{\rm SD+}$ & 135.37  & 136.42 & 136.50 & 135.41 && $124\pm 11$\\
        & $\dms$        & 141.05  & 143.11 & 145.01 & 150.00 && $147\pm 13$\\ 
 expected &$\dms$  &140.74  &140.91 &145.09   & 145.61  & & -- \\
\hline
\pb   & $m_0^{\rm SD-}$ &  1176.40  & 1188.38 & 1210.90 & 1204.92 &&--\\
        & $m_0^{\rm SD+}$ & 170.45  & 150.75 & 165.46 & 148.45 &&--\\
        & $\dms$  & 1005.95 & 1037.63 & 1045.44 & 1056.47 & & --\\ 
  expected & $\dms$  & 1013.98 & 1018.47 & 1050.14 & 1057.90 & & -- \\ 
\hline
\hline
 \end{tabular}
\caption{ Sum rule values, in fm$^2$ 
for SD excitations in \caIIp, \zr and \pb nuclei.  
The $\dms$ values in the columns
labelled with the force name are obtained as the difference
between  $m_0^{\rm SD-}$ and $m_0^{\rm SD+}$.
The  expected $\dms$ values have been calculated by using 
Eq. (\ref{eq:srSD}). 
The experimental data for \zr are taken from Ref. \cite{yak06}. 
  The values of the maximum excitation energies 
  used to integrate the RPA responses are the same as those 
  indicated in the caption of table \ref{tab:IASsr}.
}
\label{tab:SDsr}
\end{center}
\end{table}

The consistency of our calculations can be verified by observing the
results shown in Table \ref{tab:SDsr}, i.e.  the values of the,
positive and negative, zero-th energy moments (\ref{eq:mom}) and of the
SD sum rules.  These last values must be compared with those shown in
Table \ref{tab:radii} which have been obtained by using the expression
(\ref{eq:srSD}). We observe a maximal deviation of about 5\%.  In the
case of the \zr nucleus we show the experimental values of 
Ref. \cite{yak06}, and we observe the general good agreement of our
calculations within the range of the experimental uncertainties.

There are common characteristics related to the results shown in
Fig. \ref{fig:sd}.  In all the cases considered,
the strength of the $0^-$ excitations is smaller than those of the
$1^-$ and $2^-$ modes which are of similar size. Furthermore, the main
peaks of the $0^-$ responses are located at higher energies 
than the peaks of the $1^-$ and $2^-$ which almost overlap. 

Also in these charge-exchange excitations the $0^-$ state is extremely
sensitive to the tensor force as it has been observed for the charge
conserving case \cite{blo68,ang11}.  As seen in panels (a), (e) and
(i) of Fig. \ref{fig:sd}, the tensor force shifts at higher energies
the strength of this excitation mode.  The inclusion of the tensor
term is even worsening the agreement with the experimental strength
distribution disentangled in the data of Refs. \cite{wak10,wak12} for
\pb (see panel (i)).

At variance with the large effects on the $0^-$ excitations, the
tensor term does not remarkably modifies the strength distributions of
the other two SD resonances. Since the strengths of these resonances
are larger than those of the $0^-$ excitations, the total response is
scarcely affected by the presence of the tensor force (see panels (d),
(h) and (l)).  The size of these effects can be estimated by the small
changes in the centroid energies shown in Table
\ref{tab:centroid}. The general trend of these results is analogous to
that of the results of Ref. \cite{bai10}, even though the size of the
effects is smaller.

In the second part of the section we analyze in more detail the role
of the tensor force which affects our model in the HF calculations,
where it modifies the s.p. wave functions and energies which are input
of the RPA, and directly in the RPA.  In order to disentangle these
two effects, we have carried out HF and RPA calculations by switching
on and off the tensor terms of the interaction.  We label as [0,0] the
results obtained without tensor force in both HF and RPA
calculations. These do not correspond to the results obtained with the
D1M and D1S forces previously presented, since the values of the
spin-orbit terms of the D1MT2c and D1ST2c interactions are used.
We label [1,0] the results obtained by using the tensor force in HF
calculations only, and [1,1] those where the tensor force has been
used in both HF and RPA calculations. These last results are those
previously shown.  In addition, since our tensor interaction contains
two terms, see Eq. (\ref{eq:tensor2}), we have investigated separately
their relevance in the RPA calculations.  We have labelled our results
[1,t] or [1,ti] if only the pure tensor or the tensor-isospin
terms, respectively, are included in the RPA calculations. In these
two cases, the complete tensor interaction is considered in HF.

\begin{figure}[t]
\begin{center}
\includegraphics[scale=0.4] {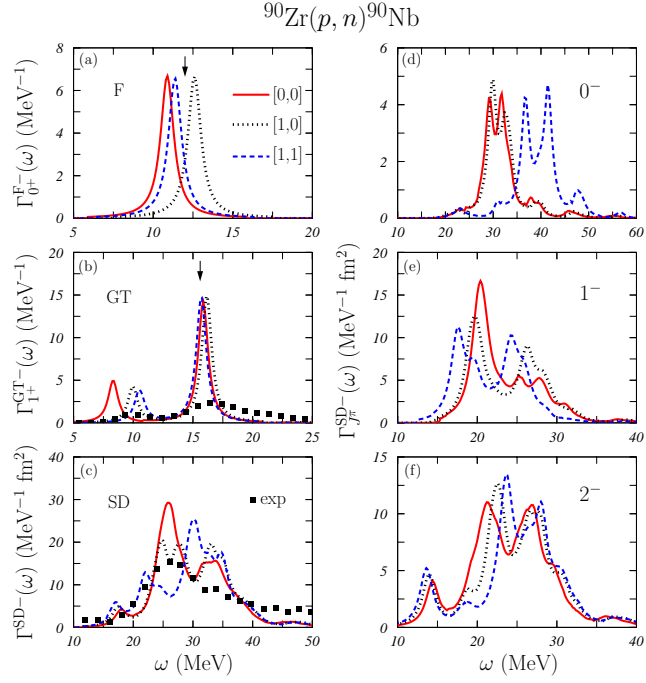}
\caption{\small (color on line) 
Energy distributions of the
\tm responses of \zr calculated with the 
D1MT2c interaction. We show
in panels (a), (b) and (c) the 
strengths $\Gamma_{0^+}^{{\rm F}-}(\omega)$, 
$\Gamma_{1^+}^{{\rm GT}-}(\omega)$ and $\Gamma^{{\rm SD}-}(\omega)$, 
respectively. The SD strengths for the 0$^-$, 1$^-$ and
2$^-$ multipoles are shown in the right panels.
The red solid lines labeled [0,0] indicate
the results obtained without tensor terms 
in both HF and RPA calculations; those labeled
[1,0] (black dotted curves) the results
obtained by considering the tensor force only in the HF calculations,
and, finally, those identified with [1,1] (blue dashed) have been obtained by
considering the tensor force in both HF and RPA calculations. 
The arrows indicate the experimental energies of the main 
peaks for the SD (Ref.\cite{bai80}) and GT (Refs. \cite{wak97,wak05})
excitations. 
Experimental data (black solid squares) are from  Ref.~\cite{wak05} 
for GT strength and from Ref.~\cite{yak06} for SD one.
}
\label{fig:t1}
\end{center}
\end{figure}

\begin{figure}[t]
\begin{center}
\includegraphics[scale=0.4] {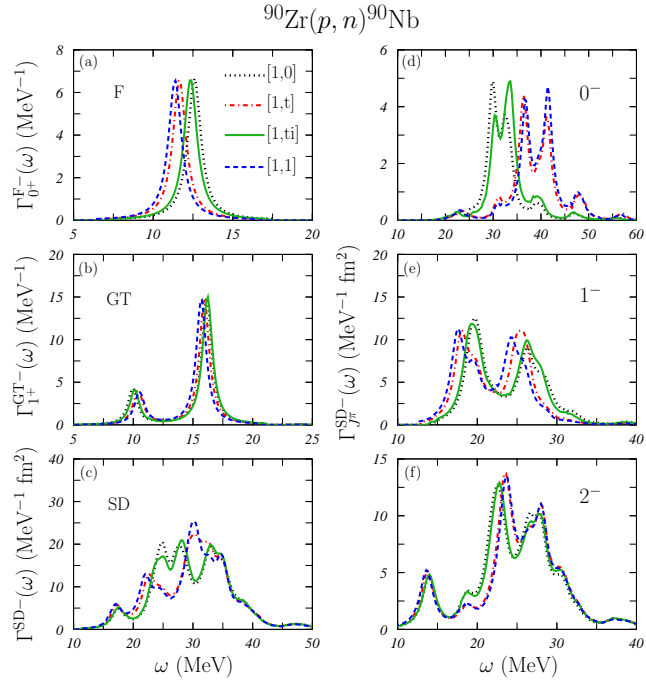}
\caption{\small (color on line) The same as in Fig. \ref{fig:t1}. 
In this case,  
all the calculations have been carried out by including tensor force   
in the HF calculations. The [1,0] black dotted and [1,1] blue dashed 
lines are the same as those
of Fig. \ref{fig:t1}, and  have been included here to facilitate
the comparison with the other results. The [1,t] red dashed-dotted
curves have been  
obtained by including only the
pure tensor force in RPA calculations, while  
those obtained by including only the tensor-isospin force are
named [1,ti] (green solid curves). 
}
\label{fig:t2}
\end{center}
\end{figure}

We conducted this study in all the three nuclei considered up to now,
and with both the D1MT2c and D1ST2c interactions. However, since we
observed rather similar effects, we present in Figs. \ref{fig:t1} and
\ref{fig:t2} only the results obtained in \zr with the D1MT2c
interaction.  In Fig. \ref{fig:t1}, we show the \tm strength
distributions obtained for the F (panel (a)) and GT (panel (b))
transitions. The results of panel (a) indicate that the IAS is
sensitive to the changes of s.p. states and energies due to the
presence of the tensor force in HF calculations.  These modifications
move the resonance peak towards higher energies.
This is compensated by a shift in the opposite direction when the tensor force is included in
the RPA calculation.

An analogous, but much smaller, sensitivity to the effect of the
tensor channel is found in the largest peak of the GT response (see
panel (b)).  Instead, the presence of the tensor force noticeably
affects the smaller peak. Its inclusion in HF calculations generates a
remarkable shift of the peak to higher energies.  The use of the
tensor force in the RPA calculations produces a smaller shift in the
same direction.

These last results can be understood in terms of the so-called Otsuka
effect \cite{ots05} present in HF calculations with the tensor
force. This effect is the main responsible of the global tensor effect
we have observed in the GT responses. In HF calculations the tensor
force between an occupied neutron state with angular momentum
$j=l+1/2$ increases the energy of the proton $j=l+1/2$ s.p. state and
lowers that of the proton $j=l-1/2$ states. This effect decreases the
energy difference between spin-orbit partners. An analogous effect of
different sign occurs with the neutron $j=l-1/2$ states. In nuclei
where all the s.p. spin-orbit partner states are occupied the two
effects compensate.  This is not the case for \zr where the last
occupied neutron state is the $\bar{n}(1g_{9/2})$.  The inclusion of
the tensor term in HF increases the s.p. energy of the proton
$p(1g_{9/2})$ state by 1.3 MeV, and reduces that of the proton
$p(1g_{7/2})$ state by 1.7 MeV.  The low-lying peak observed in panels
(b) of Figs. \ref{fig:gt} and \ref{fig:t1} is dominated by the
$\bar{n}(1g_{9/2})-p(1g_{9/2})$ p-h transition, while the
$\bar{n}(1g_{9/2})-p(1g_{7/2})$ is the main configuration in the other
peak.  Therefore the energy of the first peak is increased, while that
of the second one is reduced and the energy difference between the
two GT peaks decreases. No additional modifications of the situation
occur, since the tensor force, in this case, has very small effects on
the RPA calculations. An analogous trend is observed in \caII where
the states involved are the proton $p(1f_{7/2})$ and $p(1f_{5/2})$
s.p. states interacting with the neutron $\bar{n}(1f_{7/2})$.  This
effect explains also the upward shift of the IAS response, (see panel
(a) of Fig. \ref{fig:t1}), when the tensor force is included.

In the right panels of Fig. \ref{fig:t1} we show the
$\Gamma_{J^\pi}^{{\rm SD}-}(\omega)$ responses for each multipole
considered and in panel (c) the total $\Gamma^{{\rm SD}-}(\omega)$
strength. The effects of the tensor force are rather small in both HF
and RPA calculations for $1^-$ and $2^-$ responses. Since these are
the dominant strengths, also the total response is practically
unaffected by the inclusion of the tensor term. Different is the
situation for the $0^-$ state. While there are not effects when the
tensor is included in HF calculations, the RPA responses show a large
energy shift.

\begin{table}[htb]
\begin{center}
\begin{tabular}{ccccccccccc}
\hline\hline
&\multicolumn{1}{c}{}&\multicolumn{4}{c}{D1MT2c}&~~~~~&\multicolumn{4}{c}{D1ST2c} \\
\cline{3-6}\cline{8-11} 
&  & [1,0][0,0] & [1,t][1,0] & [1,ti][1,0] & [1,1][1,0] && [1,0][0,0] & [1,t][1,0] & [1,ti][1,0] & [1,1][1,0]  \\
\hline
\caII & $s^{{\rm F}-}$ &  1.96& -1.07& -0.28& -1.37&& 2.12& -1.17& -0.33& -1.52 \\
& $s^{{\rm GT}-}$ &  0.97&  0.49&  0.12&  0.51&& 1.00&  0.51&  0.11&  0.52 \\
& $s^{{\rm SD}-}$   &  0.62&  0.20&  0.05&  1.15&& 0.61&  0.21&  0.04&  1.12 \\ \cline{2-11}
& $s_{0^-}^{{\rm SD}-}$ &  0.35&  8.03&  1.01&  8.21&& 0.27&  8.18&  0.75&  8.01 \\ 
& $s_{1^-}^{{\rm SD}-}$ &  0.56& -1.56& -0.38& -2.17&& 0.54& -1.54& -0.38& -2.21 \\
& $s_{2^-}^{{\rm SD}-}$ &  0.72&  1.19&  0.29&  1.35&& 0.74&  1.21&  0.23&  1.34 \\
\hline
\zr & $s^{{\rm F}-}$ &  1.79 &  -1.01 &  -0.26 &  -1.29 &&   1.91 &  -1.09 &  -0.30 &  -1.42 \\
& $s^{{\rm GT}-}$ &  0.87 &   0.44 &   0.08 &   0.50 & &  0.88 &   0.46 &   0.13 &   0.48\\
& $s^{{\rm SD}-}$   &  0.43 &   0.07 &   0.00 &   0.88 & &  0.15 &   0.11 &   0.01 &   0.73\\\cline{2-11}
& $s_{0^-}^{{\rm SD}-}$ &  0.22 &   5.85 &   1.06 &   7.30 &&   0.15 &   7.00 &   0.92 &  7.23 \\
& $s_{1^-}^{{\rm SD}-}$ &  0.42 &  -1.54 &  -0.38 &  -2.10 & &  0.39 &  -1.49 &  -0.34 &  -2.16 \\
& $s_{2^-}^{{\rm SD}-}$ &  0.48 &   0.87 &   0.09 &   0.95 & &  0.47 &   0.89 &   0.10 &   0.84 \\
\hline
\pb & $s^{{\rm F}-}$ &  0.04 & -0.03 &  0.00 & -0.03 & & 0.06 & -0.03 & -0.01 & -0.04\\
& $s^{{\rm GT}-}$ &  0.19 &  0.41 &  0.09 &  0.36 & & 0.06 &  0.40 &  0.09 &  0.33\\
& $s^{{\rm SD}-}$  &  0.20 &  0.43 &  0.06 &  0.71 & & 0.38 &  0.43 &  0.04 &  0.78\\\cline{2-11}
& $s_{0^-}^{{\rm SD}-}$ &  0.32 &  7.99 &  1.56 &  9.04 &&  0.24 &  8.14 &  1.49 &  7.23\\
& $s_{1^-}^{{\rm SD}-}$ &  0.23 & -1.86 & -0.48 & -2.42 &&  0.18 & -1.81 & -0.47 & -2.40\\
& $s_{2^-}^{{\rm SD}-}$ &  0.14 &  0.80 &  0.10 &  0.92 &&  0.11 &  0.82 &  0.08 &  0.97\\
\hline
\hline
 \end{tabular}
\caption{Energy differences, in MeV, between centroid energies calculated in different manners,
as defined by Eq. (\ref{eq:cent-shift}), for all the nuclei,
interactions and transitions considered in our work.   
}
\label{tab:SDcentr}
\end{center}
\end{table}

As we have already pointed out, we obtain similar results for the
other two nuclei and for the D1ST2c interaction.  This can be seen in
Table \ref{tab:SDcentr} where we summarize the shifts between the
centroid energies obtained for different calculations by showing the
values of
\beq s^{\alpha-}([a,b][c,d]) \, = \, \omega_{\rm
  cent}^{\alpha-}[a,b] \, - \, \omega_{\rm cent}^{\alpha-}[c,d] 
  \, .
\label{eq:cent-shift}
\eeq
The letters indicated in the brackets are 0, 1, t or ti
according to the use of the tensor interaction in HF and RPA
calculations (see above).  The shifts $s_{J^\pi}^{{\rm
    SD}-}([a,b][c,d])$ calculated for the individual multipolarities
in the case of the SD transitions can be defined in an analogous way.

Coming back to the situation described above, the centroid energy in
the case of the multipolarity $0^-$ in the \tm SD transition is
shifted by 7.30 MeV in $^{90}$Zr, as indicated by the [1,1][1,0]
column for D1MT2c interaction in Table \ref{tab:SDcentr}.  In \caII
and $^{208}$Pb, $s_{0^-}^{{\rm SD}-}([1,1][1,0])$ is 8.21 and 9.04 MeV,
respectively.  These values are much larger than those found for the
other multipolarities and for the total SD transition. 
As it is shown in the
  table, we find a similar trend in all the nuclei considered and for
  both D1MT2c and D1ST2c interactions. This is a genuine effect of the
  tensor force.

For the $^{90}$Zr and the D1MT2c force, we show in Fig. \ref{fig:t2}
the comparison between the results obtained with and without the
two tensor terms in the RPA calculations.  The corresponding
shifts of the centroid energies are presented in Table
\ref{tab:SDcentr}.  These calculations have been carried out by
including the full tensor force in HF.

In general, the sensitivity to the pure tensor term is much larger
than that to the tensor-isospin one.  This can be deduced from the
fact that the results of the [1,0] (black dotted curves) and [1,ti]
(green solid curves) calculations are rather close in all the cases,
while the consideration of the pure tensor term in the RPA
calculations ([1,t] red dashed-dotted curves) produces results very
similar to the complete [1,1] calculations (blue dashed curves).  As a
consequence, the main effect of the tensor interaction is essentially
due to the presence of the pure tensor term.  An analogous effect is
observed in the F excitation.


\section{Summary and conclusions}
\label{sec:summary}

In this paper, we have presented results of charge-exchange responses
calculated within the HF+RPA framework with finite-range
interactions. These are parameter free calculations, since no part of
the interactions has been modified.
Even though we have
considered only doubly magic nuclei, however these interactions can be
used also to describe pairing effects in open shell nuclei
\cite{mar14}.  Our approach is fully self-consistent since all the
terms of the interaction used in the HF calculations have been
considered in the RPA, even the spin-orbit term which is usually
neglected in the latter calculations.

We have used two well tested parameterizations of the Gogny interaction,
the D1S and D1M forces. We have also considered two other interactions
containing tensor terms. These last interactions have been constructed
by adding to the original D1M and D1S forces a tensor and a
tensor-isospin terms. The parameters defining the tensor
force have been chosen following the procedure indicated in
Ref. \cite{gra13} which implies only a change in the strength of the
spin-orbit force.  We have investigated the role of the tensor force by
comparing the results obtained with the original Gogny forces and
those with the new interactions. 

We first remark that the HF calculations done with the original Gogny
forces generate a \caII ground state unstable for beta decay. The
interactions with tensor force stabilize the situation. 

Our study of the IAS excitations shows that the IPM is unable to
predict reasonable excitation energies, while the RPA results describe
them much better.  These excitations are characterized by a single
peak which is not very sensitive to the presence of the tensor
force. This small effect is due to a cancellation of two sizable
effects which work in opposite directions. The inclusion of the tensor
force in HF calculations changes the s.p. energies and,
consequently, we obtain peak positions which are 2 or 3 MeV
larger. When the tensor force is included in our RPA calculations we
obtain an opposite effect which essentially compensates the previous
shift.

We observe remarkable effects of the tensor force on the GT responses,
effects that help in improving the description of the experimental
data. Our GT responses are characterized by two main peaks generated
by the transitions of the excess neutrons from their last occupied
s.p. level, with orbital angular momentum $l$, to the empty proton
levels with the same orbital angular momentum.  In this case, the
major effect of the tensor force is the reduction of the energy difference
between the proton s.p. levels in HF calculations.  This implies a
reduction of the difference between the energies of the two GT peaks,
since the tensor force does not produce relevant effects in the RPA
level.

The results obtained for the SD excitation indicate that only the
$0^-$ multipole is very sensitive to the presence of the tensor force,
and this happens essentially in the RPA calculation, contrary to what
we have observed for the GT resonances. In \pb it has been possible to
disentangle the experimental SD strength attributed to the $0^-$
excitation, and we found that the inclusion of the tensor term is even
worsening the agreement with the experimental data.  We observe that
the tensor effects on the SD multipole decrease with increasing value
of the angular momentum. This behavior is similar to that found in
Ref. \cite{bai10}.

Our investigation indicates that the tensor effects we have
identified are mainly due to the pure tensor term of the interaction,
while the role of the tensor-isospin term is smaller.  

The results found in the present investigation give a reasonable
description of the experimental excitation energies but fail in
describing the width of the resonances. This is a problem related to
the intrinsic limit of the RPA which considers only 1p-1h excitations,
and it is present also in the description of charge conserving
excitations. The inclusion of two particle-two hole excitations
\cite{dro90,kam04}, or particle-vibration coupling \cite{col94}
improves the agreement with the experimental strength distributions.

We have studied the validity of our model by comparing our results
with the experimental data and with the results of other
self-consistent approaches. We conclude that the accuracy of modern
data needs the use of an effective nucleon-nucleon interaction which
contains tensor terms. A better tuning of these terms of the effective
interaction is required to apply our model in experimentally unknown
regions of the nuclear chart.

\appendix
\section{Transition matrix elements}
\label{sec:appb}

The s.p. transition operators defined in Eq. (\ref{eq:fgen})
can be expressed as the product of a term depending only on the radial
coordinate of the $i$-th particle, another term depending 
on the angular coordinates and the spin of this 
particle and a third, isospin dependent, term $\tau_\pm(i)=2t_\pm(i)$:
\beq
{\eta}^{\alpha\pm}_ {J^\pi,M}(i) \, = \, \Phi_\alpha(r_i) \, \hdb^\alpha_{J,M}(i) \, t_\pm(i) \, .
\eeq
According to  Eqs. (\ref{eq:fgen}) and (\ref{eq:f})-(\ref{eq:sd}) we have
\beqn
\Phi_{\rm F}(r_i) & = & 1 \, , \\
\Phi_{\rm GT}(r_i) & = & 1 \, , \\
\Phi_{\rm SQ}(r_i) & = & r_i^2 \, , \\
\Phi_{\rm SD}(r_i) & = & r_i \, , 
\eeqn
and
\beqn
\hdb^{\rm F}_{0,0}(i) & = & 1 \, , \\
\hdb^{\rm GT}_{1,M}(i) & = & \sqrt{4\pi} \,  [Y_0(i) \otimes \bsigma(i) ]^1_M \, ,  \\
\hdb^{\rm SQ}_{1,M}(i) & = & [Y_2(i) \otimes \bsigma(i) ]^1_M \, , \\
\hdb^{\rm SD}_{J,M}(i) & = & [Y_1(i) \otimes \bsigma(i) ]^J_M  \, .
\eeqn
 
Then, the reduced s.p.  matrix elements of Eq. (\ref{eq:strength}) can be written as
\beq
\langle a \| {\eta}^{\alpha\pm}_{J^\pi} \| b \rangle \,=\,\half 
\int {\rm d}r_i \, r_i^2 \, R^*_{a}(r_i) \, R_b(r_i) \, \Phi_\alpha(r_i)  \,
\langle l_a \half j_a \|  \hdb^\alpha_{J}(i) \|
l_b \half j_b \rangle \, \langle \half t_a   | \tau_\pm | \half t_b \rangle
\label{ed:emat2}
\, ,
\eeq
where $R_a(r)$ and $R_b(r)$ indicate
the radial part of the s.p. wave functions, 
$l_a$ and  $l_b$ are the orbital angular momenta 
of the s.p. states and $t_a$ and $t_b$  are 
the third components of their isospin.
In the above equation we have dropped the dependence
on $M$ since we have already applied the Wigner-Eckart
theorem. 

For the F operator, we have 
\beq
\langle l_a \half j_a \|  \hdb^{\rm F}_{0}(i) \| l_b \half j_b \rangle
 \, = \,  \hat{j}_a \, \delta_{l_a,l_b} \, \delta_{j_a,j_b} 
\, ,
\eeq
where we have used the symbol $\hat{j}_a = \sqrt{2j_a+1}$.
For the other operators, in case of natural parity excitations, implying $L=J$,
we can write
\beq
\langle l_a \half j_a \|  [Y_L(i) \otimes \bsigma(i) ]^J \| l_b \half j_b \rangle \, = \, 
(-1)^{l_a} \, \xi(l_a+l_b+J) \, \frac{\hat{j}_a \, \hat{j}_b
 \, \hat{J}}{\sqrt{4\pi}} \,
\left(\begin{array}{ccc}j_a &j_b &J \\ 1/2
  & 1/2 &-1  
\end{array}\right) \, .
\eeq
For unnatural parity excitations, with $L=J+s$ with $s=\pm 1$, we write
\beqn
\langle l_a \half j_a \|  [Y_L(i) \otimes \bsigma(i) ]^J \| l_b \half j_b \rangle & = & 
(-1)^{l_a+l_b+j_b+\half} \, \xi(l_a+l_b+J+1) \, \frac{\hat{j}_a \hat{j}_b}{\sqrt{4\pi}} \nonumber \\
&& \frac{\chi_a + \chi_b +sJ+\delta(s,1)}
{\sqrt{J+\delta(s,1)}} \,
\left(\begin{array} {ccc} j_a & j_b & J \\ 1/2 & -1/2 & 0 \end{array}
\right) \, ,
\eeqn
where $\xi(n)=1$ or 0 if $n$ is even or odd, 
respectively, and $\chi_a=(l_a-j_a)(2j_a+1)$.

Finally, the isospin matrix element of
Eq. (\ref{ed:emat2}) is given by
\beqn
\langle \half t_a   | \tau_+ | \half t_b \rangle &=& \delta_{a,n} \, \delta_{b,p} \, , \\
\langle \half t_a   | \tau_- | \half t_b \rangle &=& \delta_{a,p} \, \delta_{b,n} \, .
\eeqn

\acknowledgments 
This work has been partially supported by 
 the Junta de Andaluc\'{\i}a (FQM0220) and European
Regional Development Fund (ERDF) and the Spanish Ministerio de
Econom\'{\i}a y Competitividad (FPA2012-31993).


%

\begin{thebibliography}{10}
\expandafter\ifx\csname url\endcsname\relax
  \def\url#1{\texttt{#1}}\fi
\expandafter\ifx\csname urlprefix\endcsname\relax\def\urlprefix{URL }\fi

\bibitem{ost92}
F.~Osterfeld, Rev. \ Mod. \ Phys. 64 (1992) 491.

\bibitem{ich06}
M.~Ichimura, H.~Sakai, T.~Wakasa, Prog. \ Part. \ Nucl. \ Phys. 56 (2006) 446.

\bibitem{fuj11}
Y.~Fujita, B.~Rubio, W.~Gelletly, Prog. \ Part. \ Nucl. \ Phys. 66 (2011) 549.

\bibitem{arn07}
M.~Arnould, S.~Goriely, K.~Takahashi, Phys. \ Rep. 450 (2007) 97.

\bibitem{spe91}
J.~Speth, J.~Wambach, Theory of giant resonances. in Electric and magnetic
  giant resonances in nuclei, {\rm J. Speth ed.}, World Scientific, Singapore,
  1991.

\bibitem{hal67}
J.~A. Halbleib, R.~A. Sorensen, Nucl. \ Phys. \ A 98 (1967) 542.

\bibitem{lan80}
A.~M. Lane, J.~Martorell, Ann. \ Phys. \ (N.Y.) 129 (1980) 273.

\bibitem{aue81}
N.~Auerbach, A.~Klein, N.~V. Giai, Phys. \ Lett. \ B 106 (1981) 347.

\bibitem{aue83}
N.~Auerbach, A.~Klein, Nucl. \ Phys. \ A 395 (1983) 77.

\bibitem{ham00}
I.~Hamamoto, H.~Sagawa, Phys. \ Rev. \ C 62 (2000) 024319.

\bibitem{fra07}
S.~Fracasso, G.~Col\`o, Phys. \ Rev. \ C 76 (2007) 044307.

\bibitem{bai09a}
C.~L. Bai, H.~Sagawa, H.~Q. Zhang, X.~Z. Zhang, G.~Col\`o, F.~R. Xu, Phys. \
  Lett. \ B 675 (2009) 28.

\bibitem{bai09b}
C.~L. Bai, H.~Q. Zhang, X.~Z. Zhang, F.~R. Xu, H.~Sagawa, G.~Col\`o, Phys. \
  Rev. \ C 79 (2009) 041301(R).

\bibitem{bai10}
C.~L. Bai, H.~Q. Zhang, H.~Sagawa, X.~Z. Zhang, G.~Col\`o, F.~R. Xu, Phys. \
  Rev. \ Lett. 105 (2010) 072501.

\bibitem{bai11a}
C.~L. Bai, et~al., Phys. \ Rev. \ C 83 (2011) 054316.

\bibitem{bai11b}
C.~L. Bai, H.~Sagawa, G.~Col\`o, H.~Q. Zhang, X.~Z. Zhang, Phys. \ Rev. \ C 84
  (2011) 044329.

\bibitem{min13}
F.~Minato, C.~L. Bai, Phys. \ Rev. \ Lett. 110 (2013) 122501.

\bibitem{bes12}
D.~R. Bes, O.~Civitarese, J.~Suhonen, Phys. \ Rev. \ C 86 (2012) 024314.

\bibitem{civ14}
O.~Civitarese, J.~Suhonen, Phys. \ Rev. \ C 89 (2014) 044319.

\bibitem{ber91}
J.~F. Berger, M.~Girod, D.~Gogny, Comp. \ Phys. \ Commun. 63 (1991) 365.

\bibitem{gor09}
S.~Goriely, S.~Hilaire, M.~Girod, S.~P\'eru, Phys. \ Rev. \ Lett. 102 (2009)
  242501.

\bibitem{gra13}
M.~Grasso, M.~Anguiano, Phys. \ Rev. \ C 88 (2013) 054328.

\bibitem{edm57}
A.~R. Edmonds, Angular momentum in quantum mechanics, Princeton University
  Press, Princeton, 1957.

\bibitem{gaa80}
C.~Gaarde, J.~S. Larsen, M.~N. Harakeh, S.~V. van~der Werf, M.~Igarashi,
  A.~M{\"u}ller-Arnke, Nucl. \ Phys. \ A 334 (1980) 248.

\bibitem{ang12}
M.~Anguiano, M.~Grasso, G.~Co', V.~De~Donno, A.~M. Lallena, Phys. \ Rev. \ C 86
  (2012) 054302.

\bibitem{oni78}
N.~Onishi, J.~W. Negele, Nucl. \ Phys. A 301 (1978) 336.

\bibitem{firestone}
R.~B. Firestone, \url{http://isotopes.lbl.gov/toi.html}.

\bibitem{wir95}
R.~B. Wiringa, V.~G.~J. Stoks, R.~Schiavilla, Phys.\ Rev. \ C 51 (1995) 38.

\bibitem{co98b}
G.~Co', A.~M. Lallena, Nuovo \ Cimento \ A 111 (1998) 527.

\bibitem{bau99}
A.~R. Bautista, G.~Co', A.~M. Lallena, Nuovo \ Cimento \ A 112 (1999) 1117.

\bibitem{don14}
V.~De~Donno, G.~Co', M.~Anguiano, A.~M. Lallena, Phys. \ Rev. \ C 89 (2014)
  014309.

\bibitem{don09}
V.~De~Donno, G.~Co', C.~Maieron, M.~Anguiano, A.~M. Lallena, M.~Moreno-Torres,
  Phys. \ Rev. \ C 79 (2009) 044311.

\bibitem{aud03}
G.~Audi, A.~H. Wapstra, C.~Thibault, Nucl. \ Phys. \ A 729 (2003) 337.

\bibitem{audw}
http://ie.lbl.gov/toi2003/masssearch.asp.

\bibitem{vri87}
H.~de~Vries, C.~de~Jagger, C.~de~Vries, At. \ Data \ Nucl. \ Data \ Tables 36
  (1987) 495.

\bibitem{abr12}
S.~Abrahamayn, et.~al. P.R.E.X.~coll., Phys. \ Rev. \ Lett. 108 (2012) 112502.

\bibitem{ray78}
L.~Ray, G.~W. Hoffmann, G.~S. Blanpied, W.~R. Coker, R.~P. Liljestrand, Phys. \
  Rev. \ C 18 (1978) 1756.

\bibitem{yak06}
K.~Yako, H.~Sagawa, H.~Sakai, Phys. \ Rev. \ C 74 (2006) 051303(R).

\bibitem{yak09}
K.~Yako, et al., Phys. \ Rev. \ Lett. 103 (2009) 012503.

\bibitem{cla03}
B.~C. Clark, L.~J. Kerr, S.~Hama, Phys. \ Rev. \ C 67 (2003) 054605.

\bibitem{kra91}
A.~Krasznahorkay, et al., Phys. \ Rev. \ Lett. 66 (1991) 1287.

\bibitem{cha07t}
F.~Chappert, Nouvelles param\'etrisation de l'interaction nucl\'eaire effective
  de gogny, Ph.D. thesis, Universit\'e de Paris-Sud XI (France),
  http://tel.archives-ouvertes.fr/tel-001777379/en/ (2007).

\bibitem{bai80}
D.~E. Bainum, J.~Rapaport, C.~D. Goodman, D.~J. Horen, C.~C. Foster, M.~B.
  Greenfield, C.~A. Goulding, Phys. \ Rev. \ Lett. 44 (1980) 1751.

\bibitem{and85}
B.~D. Anderson, et al., Phys. \ Rev. \ C 31 (1985) 1161.

\bibitem{wak97}
T.~Wakasa, et al., Phys. \ Rev. \ C 55 (1997) 2909.

\bibitem{aki95}
H.~Akimune, et al., Phys. \ Rev. \ C 52 (1995) 604.

\bibitem{wak05}
T.~Wakasa, M.~Ichimura, H.~Sakai, Phys. \ Rev. \ C 72 (2005) 067303(R).

\bibitem{wak10}
T.~Wakasa, arxiv:1004.5220[nucl-ex].

\bibitem{wak12}
T.~Wakasa, et al., Phys. \ Rev. \ C 85 (2012) 064606.

\bibitem{col94}
G.~Col\`o, N.~Van Giai, P.~F. Bortignon, R.~A. Broglia, Phys. \ Rev. \ C 50
  (1994) 1496.

\bibitem{boh75}
A.~Bohr, B.~R. Mottelson, Nuclear structure, vol. II, Benjamin, New York, 1975.

\bibitem{conde92}
H.~Cond\'e, et al., Nucl. Phys. A 545 (1992) 785.

\bibitem{bai09}
C.~L. Bai, et~al., Phys. \ Lett. \ B 675 (2009) 28.

\bibitem{sag14}
H.~Sagawa, G.~Col\`o, arxiv:1401.6691 [nucl-th].

\bibitem{blo68}
J.~Blomqvist, A.~Molinari, Nucl. \ Phys. \ A 106 (1968) 545.

\bibitem{ang11}
M.~Anguiano, G.~Co', V.~De~Donno, A.~M. Lallena, Phys. \ Rev. \ C 83 (2011)
  064306.

\bibitem{ots05}
T.~Otsuka, T.~Suzuki, R.~Fujimoto, H.~Grawe, Y.~Akaishi, Phys. \ Rev. \ Lett.
  95 (2005) 232502.

\bibitem{mar14}
M.~Martini, S.~P\'eru, S.~Goriely, arxiv:1404.1493 [nucl-th].

\bibitem{dro90}
S.~Dro\.zd\.z, S.~Nishizaki, J.~Speth, J.~Wambach, Phys. \ Rep. 197 (1990) 1.

\bibitem{kam04}
S.~Kamerdzhiev, J.~Speth, G.~Tertychny, Phys. \ Rep. 393 (2004) 1.

\end{thebibliography}
\end{document}